\documentclass[fleqn,10pt]{wlscirep}
\usepackage[utf8]{inputenc}
\usepackage{textcomp}
\usepackage[T1]{fontenc}
\usepackage{lineno}
\usepackage{url}
\usepackage{hyperref}
\usepackage{breakurl}
\usepackage{multirow}
\usepackage{subfig}
\usepackage{float}
\usepackage{xcolor}

\usepackage[acronym]{glossaries}

\makeglossaries
\newacronym[plural=WSIs,firstplural=Whole-Slide Images (WSIs)]{wsi}{WSI}{Whole-Slide Image}
\newacronym[plural=RoIs,firstplural=Regions of Interest (RoIs)]{roi}{RoI}{Region Of Interest}
\newacronym{dl}{DL}{Deep Learning}
\newacronym{ml}{ML}{Machine Learning}
\newacronym{ai}{AI}{Artificial Intelligence}
\newacronym{he}{H\&E}{Hematoxylin \& Eosin}
\newacronym{bracs}{BRACS}{BReAst Carcinoma Subtyping}
\newacronym{cad}{CAD}{Computer-Aided-Diagnosis}

\newcommand{\ie}{\textit{i}.\textit{e}., }
\newcommand{\eg}{\textit{e}.\textit{g}., }

\title{BRACS: A Dataset for BReAst Carcinoma Subtyping in H\&E Histology Images}

\author[1*]{Nadia Brancati}
\author[2]{Anna Maria Anniciello}
\author[3,5]{Pushpak Pati}
\author[1,4]{Daniel Riccio}
\author[2]{Giosuè Scognamiglio}
\author[3,6]{Guillaume Jaume}
\author[1]{Giuseppe De Pietro}
\author[2]{Maurizio Di Bonito}
\author[3]{Antonio Foncubierta}
\author[2]{Gerardo Botti}
\author[3]{Maria Gabrani}
\author[2]{Florinda Feroce}
\author[1]{Maria Frucci}
\affil[1]{Institute for High Performance Computing and Networking of the Research Council of Italy, ICAR-CNR, Naples, Italy}
\affil[2]{National Cancer Institute -- IRCCS -- Fondazione Pascale, Naples, Italy}
\affil[3]{IBM Research -- Zurich, Switzerland}
\affil[4]{University of Naples Federico II, Naples, Italy}
\affil[5]{ETH Zurich, Switzerland}
\affil[6]{EPFL Lausanne, Switzerland}
\affil[*]{Corresponding author: nadia.brancati@cnr.it}

\begin{abstract}
Breast cancer is the most commonly diagnosed cancer and registers the highest number of deaths for women with cancer.
Recent advancements in diagnostic activities combined with large-scale screening policies have significantly lowered the mortality rates for breast cancer patients.
However, the manual inspection of tissue slides by pathologists is cumbersome, time-consuming, and is subject to significant inter- and intra-observer variability. 
Recently, the advent of whole-slide scanning systems have empowered the rapid digitization of pathology slides, and enabled to develop digital workflows. These progress further enable to leverage \gls{ai} to assist, automate, and augment pathological diagnosis.
But \gls{ai} techniques, especially \gls{dl}, require a large amount of high-quality \emph{annotated} data to learn from. Constructing such task-specific datasets poses several challenges, such as, data-acquisition level constrains, time-consuming and expensive annotations, and anonymization of patient information, etc.
In this paper, we introduce the \gls{bracs} dataset, a large cohort of annotated \gls{he}-stained images to advance the characterization of breast lesions.
\gls{bracs} contains 547 \glspl{wsi}, and 4539 \glspl{roi} extracted from the \glspl{wsi}. Each \gls{wsi}, and respective \glspl{roi}, are annotated by the consensus of three board-certified pathologists into different lesion categories. Specifically, \gls{bracs} includes three lesion types, \ie benign, malignant and atypical, which are further subtyped into seven categories. The included \glspl{roi} exhibit large variability in dimensions, and incorporate the usual tissue-preparation and staining artifacts to bestow a realistic breast cancer diagnosis. It is, to the best of our knowledge, the largest annotated dataset for breast cancer subtyping both at \gls{wsi}- and \gls{roi}-level.
Further, by including the understudied atypical lesions, \gls{bracs} offers an unique opportunity for leveraging \gls{ai} to better understand their characteristics.
We encourage \gls{ai} practitioners to develop and evaluate novel algorithms on the \gls{bracs} dataset to further breast cancer diagnosis and patient care. 
\end{abstract}

\begin{document}
\flushbottom
\maketitle
\thispagestyle{empty}

\section*{Background \& Summary}

Histology images contain both complex and ambiguous information, thus challenging pathologists to perform a robust, reproducible and efficient analysis.
Further, histology images are very large, which makes their analysis cumbersome and time-consuming.  
With advances in \gls{cad}, \gls{ai} techniques, especially \gls{ml} and \gls{dl},
have the potential 
to address the aforementioned bottlenecks~\cite{madabhushi2016image,tizhoosh2018artificial,srinidhi2020deep,de2021machine}. 
These techniques can identify discriminative morphological patterns 
from large datasets to diagnose histology images in a standardized and objective manner.
However, there exist several challenges in adopting such techniques in digital pathology, such as,
(i) the requirement of large annotated datasets, 
(ii) the need for sufficiently variable data to set up cross-patient experiments,
(iii) the inclusion of diagnostically challenging lesions, that are generally difficult and expensive to acquire,
(iv) the utilization of sub-region annotations to delineate \gls{roi},
(v) the coverage of diagnostic spectrum, and
(vi) coping with data leakage and noisy annotations.
Recent advancements in \gls{dl} have demonstrated superior capabilities compared to classical \gls{ml} approaches for \gls{cad}~\cite{araujo2017classification,bardou2018classification,sudharshan2019multiple,duggento2020deep,benhammou2020breakhis,sharma2020conventional,chugh2021survey}. The crucial advantage of \gls{dl} approaches is their ability to learn task-specific salient features directly from the training data.
However, this superiority comes at the cost of acquiring large, high-quality, variable, and unbiased annotated training datasets.
Although several datasets for diagnosing breast histology images exist~\cite{idc2014,spanhol2015dataset,bejnordi2017,aresta2019,veta2018tupac}, they do not meet all the aforementioned criteria. 
For instance, some datasets focus on specific diseases that include only binary classes~\cite{bejnordi2017,idc2014}, while others handling multiple classes~\cite{spanhol2015dataset,aresta2019} include only a small number of training samples (both at \gls{wsi}- and \gls{roi}-level) collected from a few patients, thus limiting the dataset variability.
Further, the set of considered classes in a dataset is crucial.
Most of the public datasets aim to categorize lesions into benign and malignant classes, which do not depict the complete spectrum of classes in breast cancer diagnosis.
Many of these datasets contain standardized images without clinical artifacts, \eg staining anomalies, ink marks, tissue folding, blurred regions, tears etc.
Consequently, these datasets do not comprehensively represent the real-world breast cancer diagnosis. 
Thus, it is necessary to develop a breast cancer dataset consisting of heterogeneous images across the diagnostic spectrum which is comparable to real-world diagnosis performed by the pathologists.

To this end, we introduce \gls{bracs}, a large cohort of \gls{he}-stained images to advance \gls{cad} of breast lesions. 
\gls{bracs} features the following advantages over the extant breast cancer image datasets,
(i) it includes a large and heterogeneous set of realistic breast histology images (both at \gls{wsi}- and \gls{roi}-level), 
(ii) \glspl{roi} range over variable dimensions by entirely including the diagnostic lesion, thus avoiding the loss of diagnostically relevant information, 
(iii) the images are acquired from a large number of patients encompassing large variability, and 
(iv) two atypical lesion categories, also known as precancerous lesions, are included along with other categories.
In particular, we consider the following lesion types, Normal (N), Pathological Benign (PB), Usual Ductal Hyperplasia (UDH), Flat Epithelial Atypia (FEA), Atypical Ductal Hyperplasia (ADH), Ductal Carcinoma in Situ (DCIS), and Invasive Carcinoma (IC). 
Thus, \gls{bracs} represents a more realistic benchmark for breast cancer diagnosis by including several types of typical and atypical tissue samples over a wide variety of \glspl{wsi} and \glspl{roi} extracted from a large number of patients.


\section*{Methods}


The \gls{bracs} dataset is created to support the development of breast cancer diagnostic methods through the automatic analysis of histology images.
The dataset was developed through the collaboration of the National Cancer Institute - Scientific Institute for Research, Hospitalization and Healthcare (IRCCS) "Fondazione G. Pascale", the Institute for High Performance Computing and Networking (ICAR) of National Research Council (CNR), and IBM Research -- Zurich.
The dataset was acquired from patients between 2019 and 2020, by board-certified pathologists of the Department of Pathology at the National Cancer Institute - IRCCS "Fondazione G. Pascale" in Naples (Italy).
The samples were generated from \gls{he}-stained breast tissue biopsy slides, and were selected based on the diagnostic reports of the patients. 
The age of the patients range from 16 to 86 years, with about 61\% of patients in the range of 40-60 years, and only a few patients aging less than 20 years or above 80 years.

\subsection*{\gls{wsi}- and \gls{roi}-level annotations}

\begin{figure*}
      \centering
      \includegraphics[width=400pt]{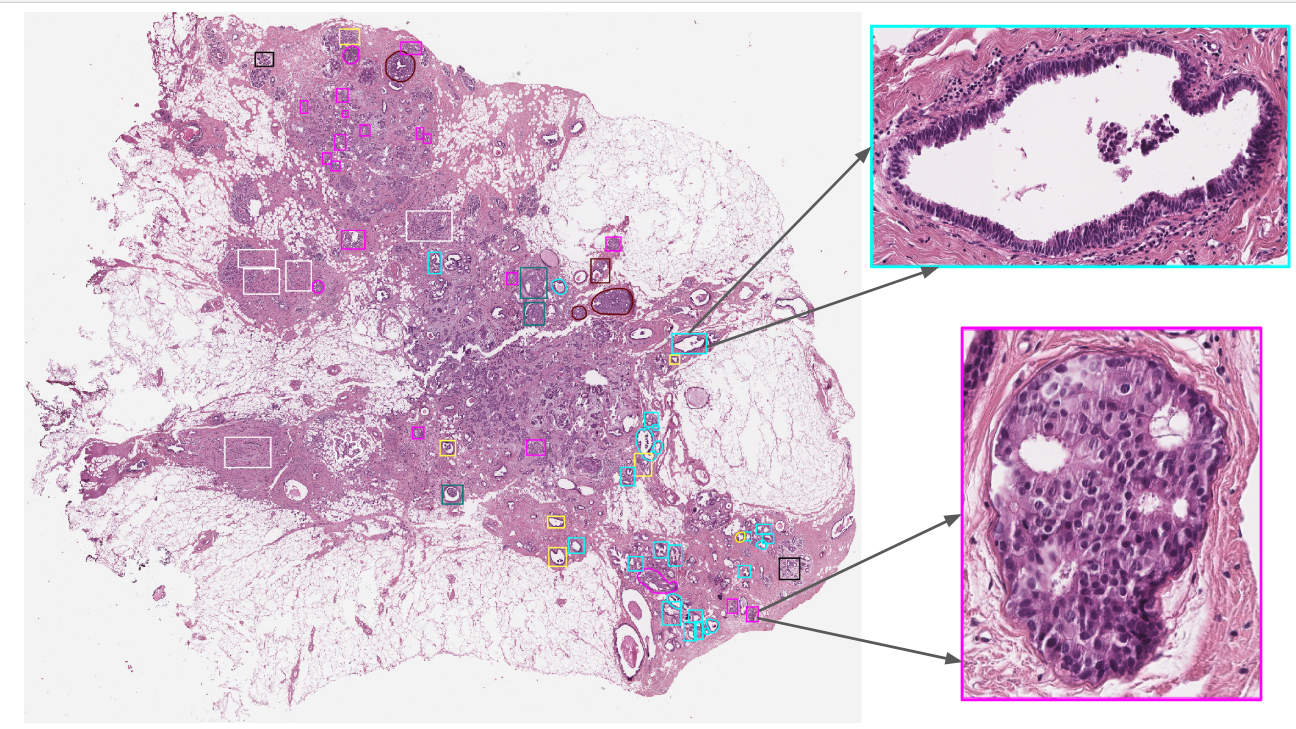}
      \caption{Example of a \gls{wsi} and its associated \glspl{roi}. The color of each contour indicates the tumor subtype of the lesion according to a fixed color palette.}
      \label{figure:WSIRoI}
\end{figure*}

The curation of rich and comprehensively annotated histology images is a complex and time-consuming task, while being prone to observer variability. The inclusion of atypical breast lesions at both \gls{wsi}- and \gls{roi}-level further increases the task complexity by requiring annotations of specialized expert pathologists.
Moreover, a \gls{wsi} typically includes \emph{several} lesions of \emph{different} subtypes.
To address the aforementioned challenges, we started by extracting and annotating \glspl{roi} in \glspl{wsi}, and subsequently derived the \gls{wsi}-level label as the most severe cancerous lesion detected within the slide. Specifically, the \gls{roi}-level annotations were conducted in a two-step procedure. 
First, a set of representative \glspl{roi} in each \gls{wsi} was identified. 
Three board-certified pathologists \emph{independently} annotated the \glspl{roi}, \ie either as normal tissue or as one of the six lesion subtypes. Each extracted \gls{roi} corresponds to a unique category, and can include single or multiple glandular structures.
Then, the annotations with disagreement were further discussed and re-annotated by the consensus of three pathologists.
This process ensures reliable annotations, and allows to alleviate inter- and intra-observer variability.
Figure~\ref{figure:WSIRoI} presents the annotation procedure for a sample \gls{wsi} and its corresponding \glspl{roi}. All the \glspl{roi} were extracted and annotated with the help of QuPath~\cite{bankhead2017}.
The number of extracted \glspl{roi} per \gls{wsi} ranges from 0 to 119, with an average of 11 \glspl{roi} per \gls{wsi}. 
Multiple \glspl{roi} corresponding to different lesion subtypes were selected per \gls{wsi} such that the collected \glspl{roi} across all \glspl{wsi} collectively encompass the lesion heterogeneity.
This aspect is very crucial for representing pathogenesis and disease progression, and consequently allowing for the inclusion of sufficiently variable data for \gls{dl} model training.
Subsequently, the \gls{wsi}-level labels were trivially defined as the most aggressive tumor subtype annotated in the image. 

\subsection*{Subtype annotations}

An important characteristics of \gls{bracs} is the inclusion of atypical lesions, ADH and FEA. While ignored in other public datasets, these categories remain important as they might be indicators of either (i) the presence of abnormalities in neighboring breast tissue that could go undetected, \eg due to the extraction of small tissue samples, or (ii) a high risk of onset of future carcinoma, \ie development of DCIS and IC. 
In addition, these lesions cannot be detected by mammography or other breast imaging techniques, nor can they be felt during a clinical breast examination. When detected in a core biopsy, more frequent imaging follow-up and often surgical excision are recommended~\cite{ingegnoli2010flat}.
\gls{bracs} also includes lesion subtypes belonging to benign and malignant types. In particular, benign lesions are subtyped as either non-cancerous lesions (PB) or inflammatory responses (UDH).
Malignant lesions are categorized as either DCIS or IC. 
Finally, histology images representing normal tissue sample are classified in the Normal (N) category.

\begin{figure*}
      \centering
      \subfloat[]
     {\includegraphics[width=125pt]{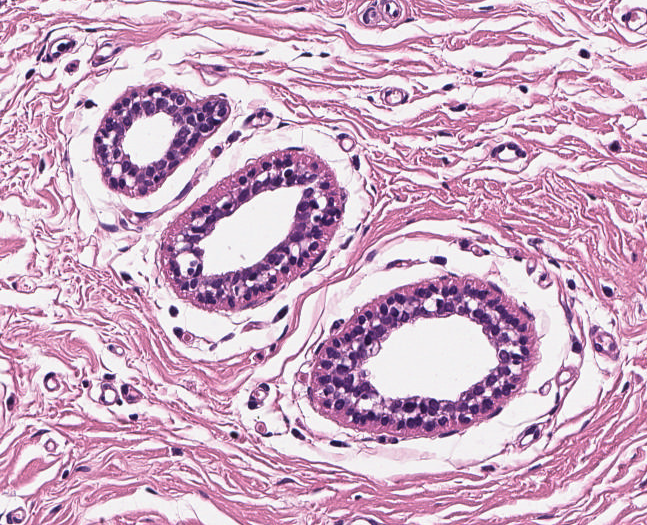}}
     \hspace{0.5mm}
     \subfloat[]
      {\includegraphics[width=113pt]{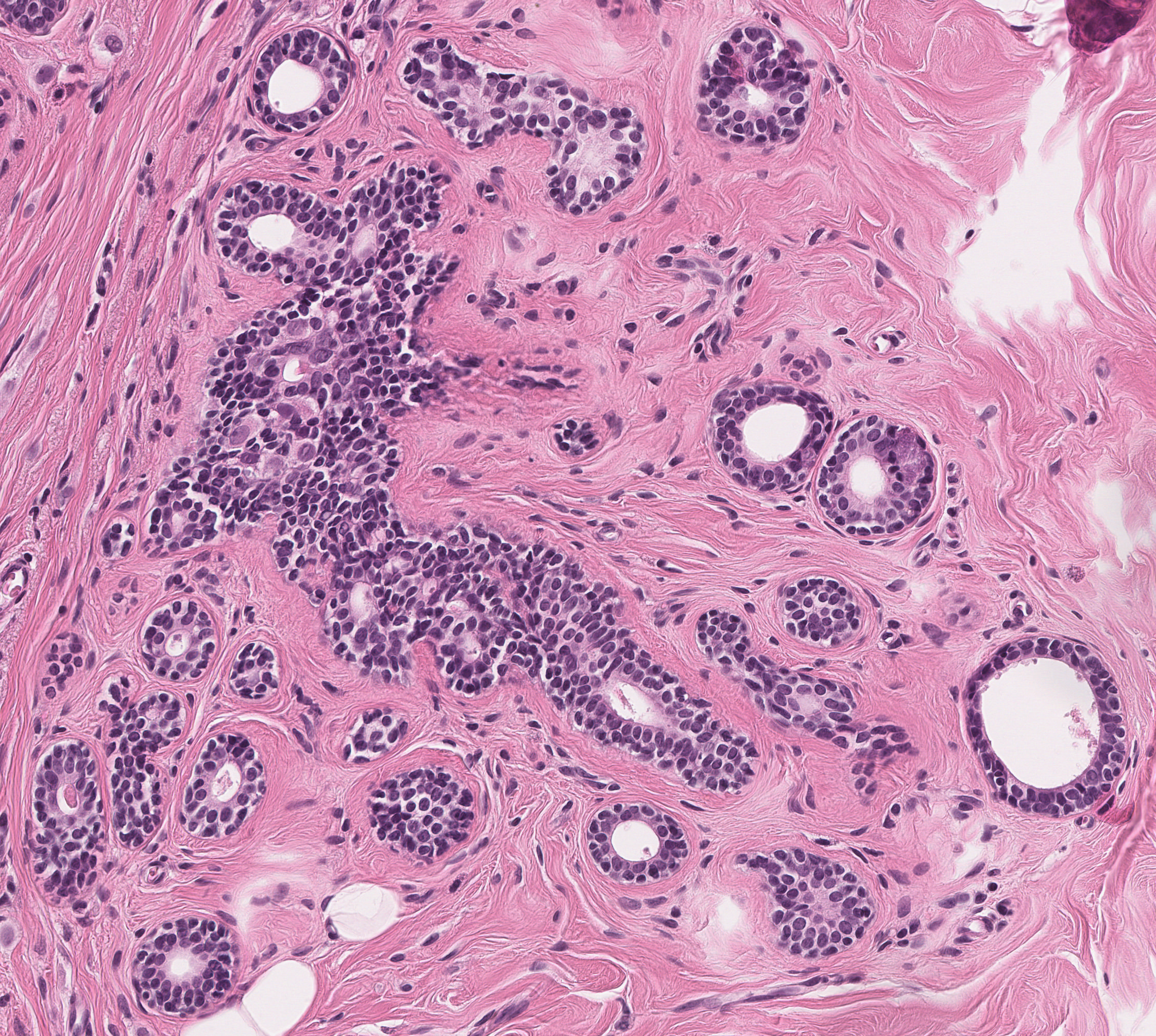}}
      \hspace{0.5mm}
      \subfloat[]
      {\includegraphics[width=164pt]{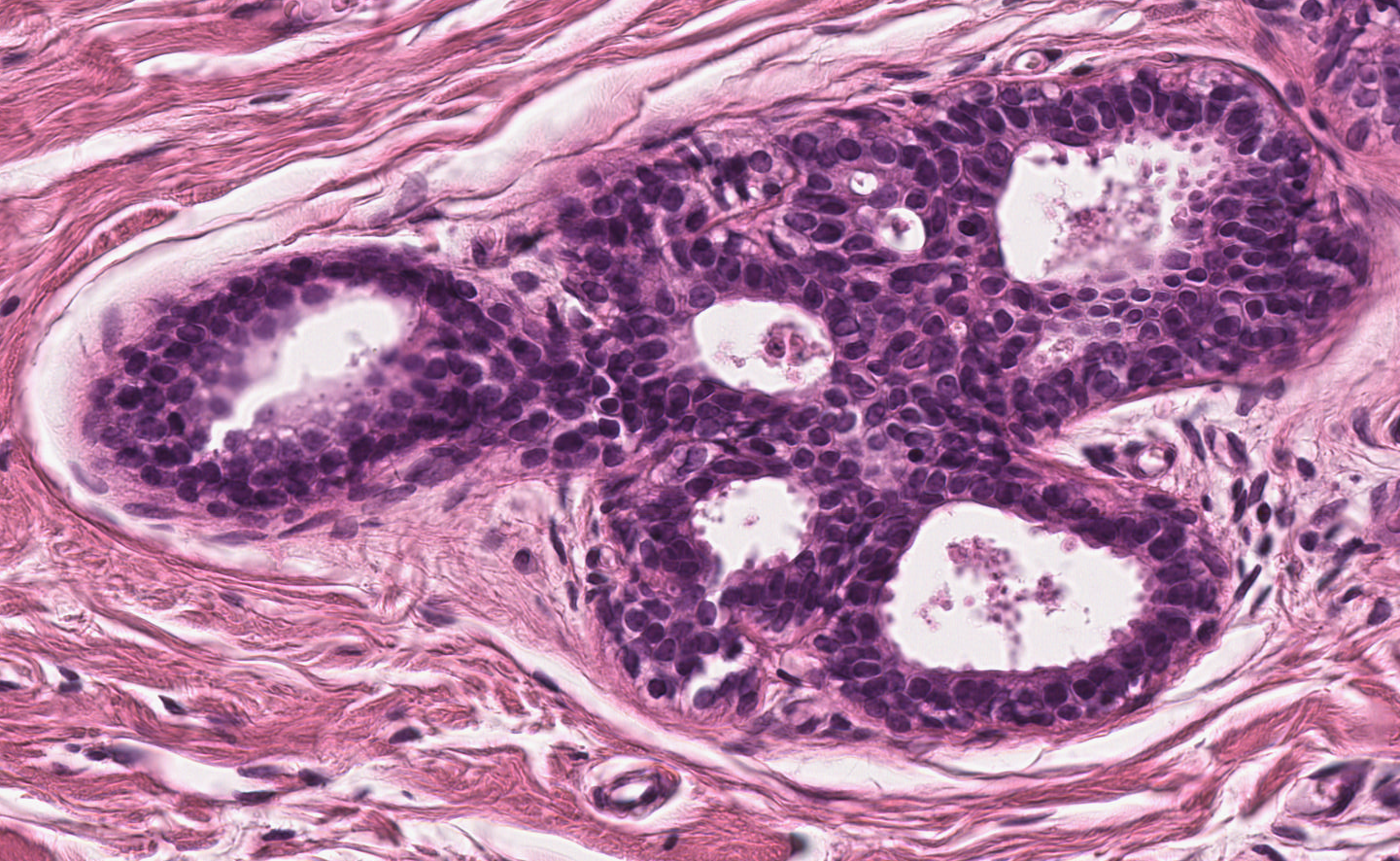}}\\
      \vspace{0.5cm}
    \subfloat[]
     {\includegraphics[width=200pt]{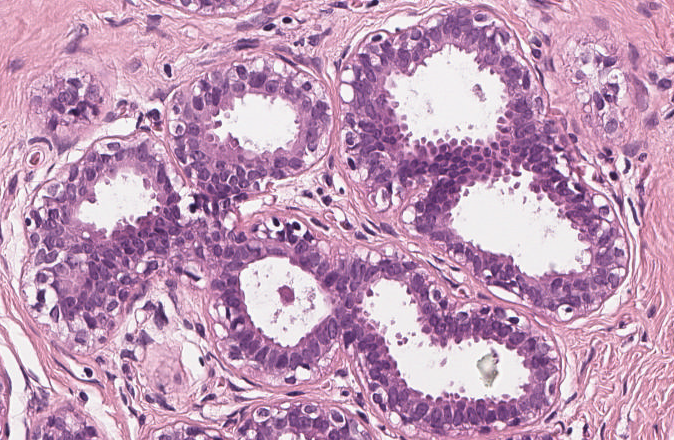}}
     \hspace{0.5mm}
     \subfloat[]
      {\includegraphics[width=126pt]{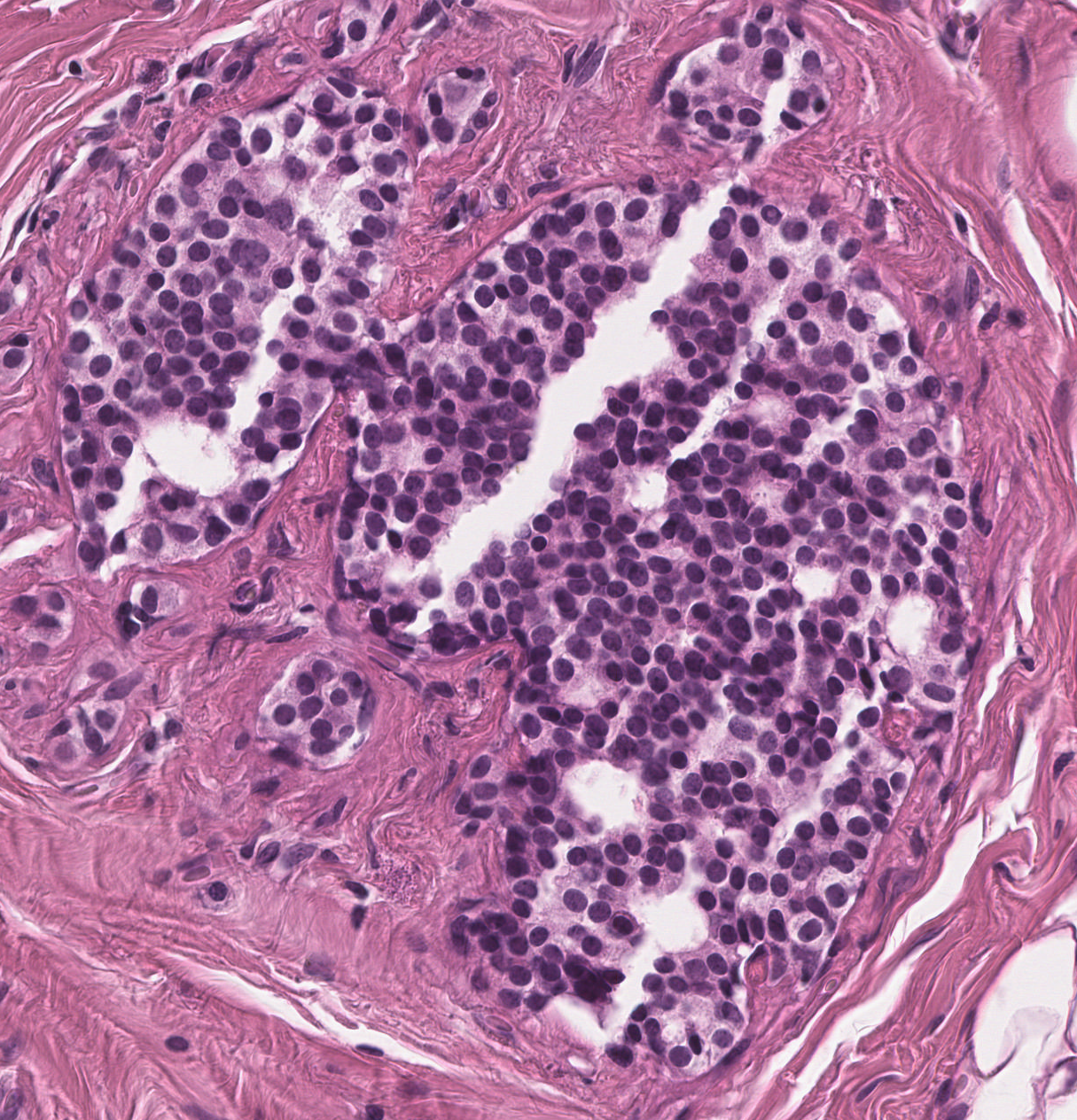}}\\
      
      \vspace{0.5cm}
    \subfloat[]
     {\includegraphics[width=184pt]{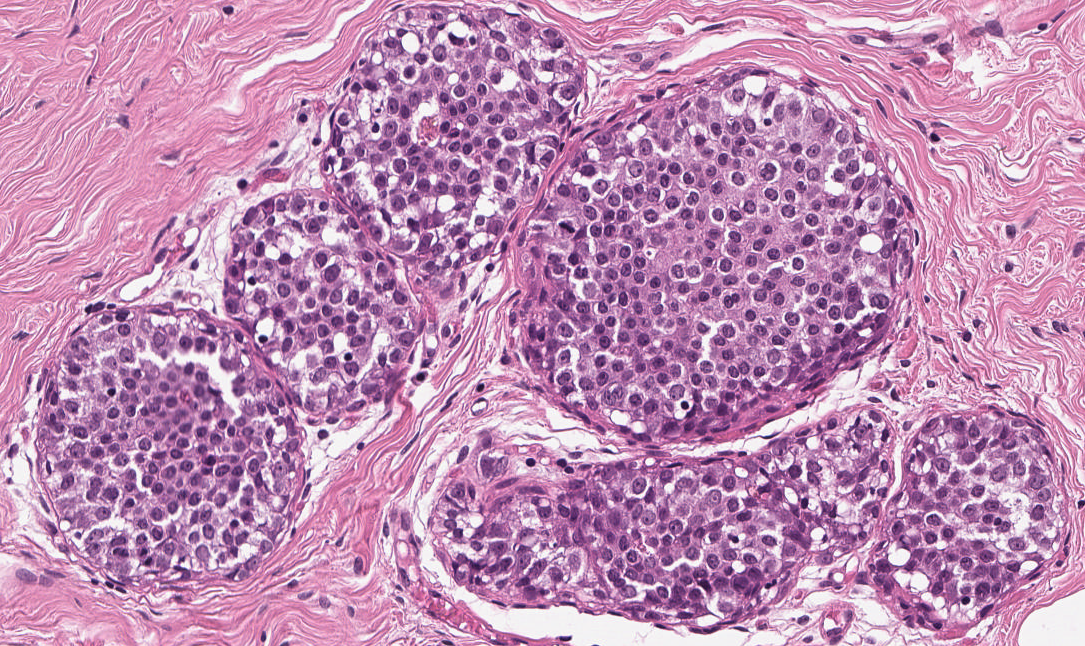}}
     \hspace{0.5mm}
     \subfloat[]
      {\includegraphics[width=174pt]{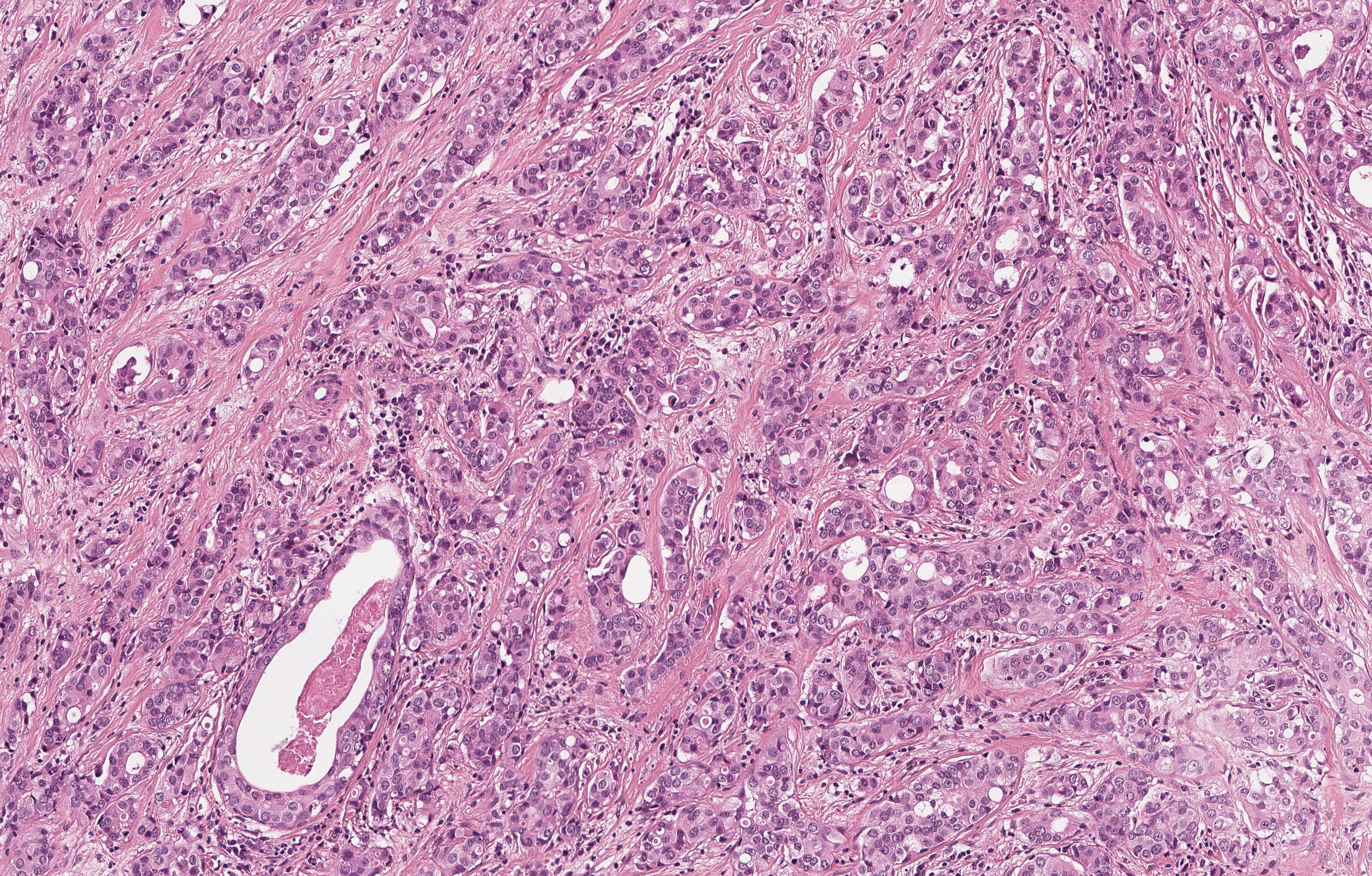}}\\
      
      \caption{Examples of different tissue samples: (a) Normal (N), (b) Pathological Benign (PB), (c) Usual Ductal Hyperplasia (UDH), (d) Flat Epithelial Atypia (FEA), (e) Atypical Ductal Hyperplasia (ADH), (f) Carcinoma in Situ (DCIS) and (g) Invasive Carcinoma (IC).}
      \label{figure:samples}
\end{figure*}

In order to clarify the description of the different tissue subtypes, a brief description of the mammary gland should be considered. The breast is a modified apocrine sweat gland, made up of 15-25 independent glandular units called lobes, each of which is formed by a compound tubulo-acinar gland. The lobes are comprised of adipose tissue and divided by connective tissue septa. Inside each lobe, the main ducts branch into terminal ducts, each of which leads to a lobule that is made up of many berries to form the ductulo-lobular terminal unit. Detailed information on lesions included in \gls{bracs} can be found in \cite{who2019}. 
The specific features of the different sample tissue subtypes are shortly summarized in the following and a representative example for each of them is shown in Figure~\ref{figure:samples}.

\subsubsection*{Normal Tissue (N)}
In normal mammary glandular tissue, there are two types of epithelial cells (the luminal layer and the basal myoepithelial layer) and two types of stromal cells (interlobular stroma and intralobular stroma). Differently from PB, the ratio between epytelial component and stroma is preserved.

\subsubsection*{Pathological Benign (PB)}
Benign breast lesions can be grouped according to the risk of developing invasive carcinoma and include several groups of histological entities classified in relation to morphology. In our study, for reasons of differential diagnosis, in the PB category we included both non-proliferative lesions and proliferative lesions with the exception of UDH, FEA and ADH, which were considered as three independent subtypes. Therefore, PB includes cyst, apocrine metaplasia, ductal ectasia, squamous metaplasia, atrophy, stromal fibrosis, mastitis, sclerosing adenosis, papilloma, radial scar, simple and complex fibroadenoma.

\subsubsection*{Usual Ductal Hyperplasia (UDH)}
UDH is characterized by an increase in the epithelial layers. It is a cohesive proliferation of disorderly distributed but oriented cells. It can have different architectural aspects (solid pattern, fenestrated pattern, micropapillary pattern). Even if UDH shares some architectural features with ADH and DCIS, it does not show atypia.

\subsubsection*{Flat Epithelial Atypia (FEA)}
FEA is a proliferative lesion characterized by low grade cytological atypia, cell monomorphism, loss of polarity and orientation with respect to the basement membrane, presence of apical snout, endoluminal secretion and frequent calcifications.

\subsubsection*{Atypical Ductal Hyperplasia (ADH)} 
ADH is a proliferation of monomorphic cells, which only partially fill the ductal spaces. Architectural aspects include a solid pattern, a cribriform pattern and a papillary pattern.
The cytologic atypia is similar to that of low-grade DCIS, but the lesion spans no more than 2mm or has an insufficient architectural atypia involving only partially ducts and / or lobules.

\subsubsection*{Ductal Carcinoma in Situ (DCIS)}
In situ carcinoma is a malignant proliferation of epithelial cells that fills the entire duct, without evidence of stroma invasion. Typically it involves multiple adjacent ductal spaces. It can have cribriform, solid, papillary and micropapillary patterns.

\subsubsection*{Invasive Carcinoma (IC)}
IC is characterized by the invasion of tumor cells infiltrating the breast stroma with loss of peripheral myoepithelial cells. The presence of the myoepithelial cell layer is an important distinction of DCIS from IC.
\\

Detecting certain subtypes is particularly challenging as some morphological patterns can be shared by several classes.
For instance, ADH shares morphological similarities with DCIS. In certain cases, it even includes all the features of DCIS, but is simply limited in size.
Also, UDH, ADH and DCIS are all characterized by an intraductal growth pattern, which makes these classes difficult to classify and differentiate in \gls{he}-stained sections.
This is particularly true when pathologists cannot utilize significant \gls{wsi}-level context.
We emphasize that when classifying \gls{bracs} \glspl{roi}, no context information is provided beside the tissue lesion and immediate tissue micro-environment.

\subsection*{\gls{bracs} dataset characteristics}

The \gls{bracs} dataset contains 547 \glspl{wsi} extracted on 189 different patients. It also includes 4'539 \glspl{roi} extracted from 387 \gls{wsi} collected on 151 patients. All slides were scanned with an Aperio AT2 scanner at $0.25$ $\mu m/pixel$ using a magnification factor of $40\times$.  Table~\ref{table:table1} and Table~\ref{table:table2} report the number of \glspl{wsi} (with and without \glspl{roi}) and \glspl{roi} according to the lesion type and subtypes, respectively.

\begin{table}[H]
\centering
	\begin{tabular}{  c | c | c | c | c } 
	 \hline  \textbf{Data} & \textbf{Benign} & \textbf{Atypical} & \textbf{Malignant} & \textbf{Total}  \\ 
	\hline \glspl{wsi} with \glspl{roi} &  $149$ & $75$ & $163$ & $387$\\
	\hline \glspl{wsi} without \glspl{roi} &  $116$ & $14$ & $30$ & $160$ \\
	\hline \glspl{wsi} &  $265$ & $89$ & $193$ & $547$ \\ 
	\hline \glspl{roi} &  $1837$ & $1263$ & $1439$ & $4539$ \\ 
    \hline
	\end{tabular}
	\caption{\gls{bracs} data distribution according to lesion type.}
	\label{table:table1}
\end{table}

\begin{table} [H]
\centering
	\begin{tabular}{  c | c | c | c | c | c | c | c } 
	\hline   \textbf{Data} & \textbf{N} & \textbf{PB} & \textbf{UDH} & \textbf{FEA} & \textbf{ADH} & \textbf{DCIS} & \textbf{IC}  \\ 
	\hline \glspl{wsi} with \glspl{roi} &  $17$ & $77$ & $55$ & $34$  & $41$  & $51$  & $112$\\
	\hline \glspl{wsi} without \glspl{roi} &  $27$ & $70$ & $19$ & $7$  & $7$  & $10$  & $20$ \\
	\hline \glspl{wsi} &  $44$ & $147$ & $74$ & $41$  & $48$  & $61$  & $132$ \\ 
	\hline \glspl{roi} &  $484$ & $836$ & $517$ & $756$  & $507$  & $790$  & $649$ \\ 
    \hline
	\end{tabular}
	\caption{\gls{bracs} data distribution according to lesion subtype.}
	\label{table:table2}
\end{table}


\subsection*{Comparison with existing datasets}

In recent years, several data sets have been proposed, with more and more samples and classes~\cite{bach2018,roux2013,bio2015,camelyon16,tupac16,camelyon17}.
Table~\ref{table:table7} details existing public datasets of histology images for breast lesion classification. 
Datasets that are (i) either no more accessible \cite{roux2013}, (ii) subsets of already mentioned datasets~\cite{bio2015}, or (iii) targeting specific tasks, \eg lesion proliferation scores prediction~\cite{tupac16}, pN-stage prediction~\cite{camelyon17}, are not mentioned in Table~\ref{table:table7}. 

The IDC~\cite{idc2014} and Camelyon16~\cite{camelyon16} datasets are focusing on the detection of the presence of a given lesion. In particular, IDC provides \glspl{roi} at small spatial resolution ($50 \times 50$ pixels) extracted from large areas of Invasive Ductal Carcinoma. We emphasize that even if the number of \glspl{roi} in \gls{bracs} is lower than in IDC, \gls{bracs} images are on average much larger allowing the inclusion of whole glandular areas. \gls{bracs} includes a larger number of \glspl{wsi} than Camelyon16 and more subtypes.
BreakHis~\cite{break2015} and BACH~\cite{bach2018} datasets are devoted to multi-classification tasks, but remain significantly smaller than \gls{bracs}, both in terms of image size and number of samples, and include less subtypes than \gls{bracs}. 

In conclusion, the number of \glspl{wsi}, \glspl{roi} and associated patients, is significantly larger in \gls{bracs} compared to the existing public datasets.
Moreover, BreakHis and BACH include fixed-size \glspl{roi}, while \gls{bracs} images are of arbitrary size. Assuming fixed-size tumor regions is a strong assumption that does not apply in real-life scenarios. 
Limiting the size of \glspl{roi} imposes to either (i) partially cut the lesion, hence producing a loss of information that could be pivotal for a correct diagnosis, or (ii) manually curating \glspl{roi} such that they all have similar sizes, which does not encompass tumor heterogeneity.
By proposing samples of varying sizes, \gls{bracs} promotes the development of \gls{dl} algorithms that need to be able to operate on inputs of different dimensionality.
\gls{bracs} is also the first dataset to include pre-cancerous atypical lesions.
\gls{bracs} and BACH share the same malignant lesion subtypes, while BreakHis refines this class by partitioning it into four specific subtypes. All the benign lesion subtypes defined in BACH and BreakHis are also included in \gls{bracs} (normal and benign).
In addition, \gls{bracs} includes UDH lesion subtype that is not considered in BACH and BreakHis.

In summary, \gls{bracs} characteristics are unique, as it allows for  multi-class classification task of breast cancer lesions, including challenging atypical lesions. \gls{bracs} is also larger than the existing datasets in terms of number \glspl{roi}, \glspl{wsi} and patients.


\begin{table}[t]
\begin{tabular}{l|l|l|l|l|l}
\hline
\multirow{2}{*}{\textbf{Dataset, Year}} &
  \multicolumn{2}{c|}{\textbf{Lesion Classes}} &
  \multirow{2}{*}{\textbf{\begin{tabular}[c]{@{}l@{}}Data Type \\ Size (Magnification)\end{tabular}}} &
  \multirow{2}{*}{\textbf{n. Pat.}} &
  \multirow{2}{*}{\textbf{\begin{tabular}[c]{@{}l@{}}Resolution \\ in pixels\end{tabular}}} \\ \cline{2-3}
 &
  \textbf{Benign} &
  \textbf{Malignant} &
   &
   &
   \\ \hline
\begin{tabular}[c]{@{}l@{}} Invasive Ductal Carcinoma \\(IDC) \cite{idc2014}, 2014\end{tabular} &
  IDC negative &
  IDC positive &
  \begin{tabular}[c]{@{}l@{}}\textit{\gls{roi}} \\ 277.524 (40x)\end{tabular} &
  162 &
  50x50 \\ \hline
BreakHis \cite{spanhol2015dataset}, 2015 &
  \begin{tabular}[c]{@{}l@{}}Adenois \\ Fibroadenoma \\ Phyllodes tumor\\ Tubular adenona\end{tabular} &
  \begin{tabular}[c]{@{}l@{}}Carcinoma\\ Lobular carcinoma \\ Mucinous carcinoma \\ Papillary carcinoma\end{tabular} &
  \begin{tabular}[c]{@{}l@{}}\textit{\gls{roi}}\\ 1.995 (40x) \\ 2.081 (100x) \\ 2.013 (200x)  \\ 1.820 (400x)\end{tabular} &
  82 &
  700x460 \\ \hline
\multirow{2}{*}{Camelyon16 \cite{camelyon16}, 2016} &
  \multirow{2}{*}{\begin{tabular}[c]{@{}l@{}}Normal\\ Benign\end{tabular}} &
  \multirow{2}{*}{\begin{tabular}[c]{@{}l@{}}In-situ carcinoma\\ invasive carcinoma\end{tabular}} &
  \multirow{2}{*}{\begin{tabular}[c]{@{}l@{}}\textit{\gls{wsi}} \\ 400 (20x and 40x)\end{tabular}} &
  \multirow{2}{*}{400} &
  \multirow{2}{*}{Variable Size} \\
 &
   &
   &
   &
   &
   \\ \hline
BACH \cite{bach2018}, 2018 &
  \begin{tabular}[c]{@{}l@{}}Normal\\ Benign\end{tabular} &
  \begin{tabular}[c]{@{}l@{}}In-situ carcinoma\\ invasive carcinoma\end{tabular} &
  \begin{tabular}[c]{@{}l@{}}\textit{\gls{roi}}\\ 400 (200x)\\ \textit{\gls{wsi}}\\ 10(20x)\end{tabular} &
  \begin{tabular}[c]{@{}l@{}}39\\ \\ 10\end{tabular} &
  \begin{tabular}[c]{@{}l@{}}2048×1536 \\ \\ Variable Size\end{tabular} \\ \hline
\end{tabular}
\caption{Popular publicly available breast histopathology image dataset.}
\label{table:table7}
\end{table}

\begin{table}[t]
\centering
\begin{tabular}{l|l|l|l|l|l|l}
\hline
\multirow{2}{*}{\textbf{Dataset, Year}} &
  \multicolumn{3}{c|}{\textbf{Lesion Classes}} &
  \multirow{2}{*}{\textbf{\begin{tabular}[c]{@{}l@{}}Data Type \\ Size (Magnification)\end{tabular}}} &
  \multirow{2}{*}{\textbf{n. Pat.}} &
  \multirow{2}{*}{\textbf{\begin{tabular}[c]{@{}l@{}}Resolution \\ in pixels\end{tabular}}} \\ \cline{2-4}
 &
  \textbf{Benign} &
  \textbf{Atypical} &
  \textbf{Malignant} &
   &
   &
   \\ \hline
\gls{bracs} \cite{bracs2021}, 2021 &
  \begin{tabular}[c]{@{}l@{}}Normal\\ Benign\\ UDH\end{tabular} &
  \begin{tabular}[c]{@{}l@{}}FEA\\ ADH\end{tabular} &
  \begin{tabular}[c]{@{}l@{}}In-situ carcinoma\\ invasive carcinoma\end{tabular} &
  \begin{tabular}[c]{@{}l@{}}\textit{\gls{roi}}\\ 4537 (40x)   \\ \textit{\gls{wsi}}\\ 547 (40x)\end{tabular} &
  \begin{tabular}[c]{@{}l@{}}151\\ \\ 189\end{tabular} &
  \begin{tabular}[c]{@{}l@{}}Variable size\\ \\ Variable size\end{tabular} \\ \hline
\end{tabular}
\caption{General information on \gls{bracs} dataset}
\label{table:table8}
\end{table}
 
\subsection*{Dataset splitting}

To foster reproducibility and following \gls{ml} best practices, we provide pre-defined \gls{wsi}- and \gls{roi}-level splits in training, validation and test sets. Data split was generated such that all the \glspl{wsi} extracted from a patient belong to the \emph{same} set. Similarly, all the \glspl{roi} extracted in a given \gls{wsi} are assigned to the \emph{same} split.
By following this approach, we avoid that different sets include correlated patient- and slide-level information, that could be leading to overly optimistic prediction results~\cite{bussola2021}.
Table~\ref{table:table3} and Table~\ref{table:table4} present the number of \glspl{wsi} included in the train, validation, and test splits. Information about the number of patients in each set is provided in Table~\ref{table:table3}. Equivalent information for \glspl{roi} is shown in Table~\ref{table:table5} and Table~\ref{table:table6}.
At \gls{roi}-level, the ratio between the most common subtype (PB with 714 samples) and the least common one (N with 357) is around two. At \gls{wsi}-level, the most common subtype is PB with 120 samples and the least common one is FEA with 24 samples, hence of ratio of around four. Considering the constraint of patient- and \gls{wsi}-level split, and the fact that atypical lesions are more rare than malignant ones, \gls{bracs} offers a rather balanced set, that can directly be used for training \gls{dl} systems. 

\begin{table} [H]
\centering
	\begin{tabular}{ l | c | c | c | c | c }
	\hline    & \textbf{Benign} & \textbf{Atypical} & \textbf{Malignant} & \textbf{Total \glspl{wsi}} & \textbf{Total Patients}   \\ 
	\hline Train &  $203$ & $52$ & $140$ & $395$  & $133$ \\
	\hline Validation &  $30$ & $14$ & $21$ & $67$  & $25$ \\
	\hline Test &  $32$ & $23$ & $32$ & $85$  & $31$ \\ 
	\hline
	\end{tabular}
	\caption{\gls{wsi}-level split according to the lesion type.}
	\label{table:table3}
\end{table}

\begin{table} [H]
\centering
	\begin{tabular}{ l | c | c | c | c | c | c | c }
	\hline    & \textbf{N} & \textbf{PB} & \textbf{UDH} & \textbf{FEA} & \textbf{ADH} & \textbf{DCIS} & \textbf{IC}  \\ 
	\hline Train &  $27$ & $120$ & $56$ & $24$  & $28$  & $40$  & $100$\\
	\hline Validation &  $10$ & $11$ & $9$ & $6$  & $8$  & $9$  & $12$ \\
	\hline Test &  $7$ & $16$ & $9$ & $11$  & $12$  & $12$  & $20$ \\ 
	\hline
	\end{tabular}
	\caption{\gls{wsi}-level split according to the lesion subtype.}
	\label{table:table4}
\end{table}

\begin{table} [H]
\centering
	\begin{tabular}{l | c | c | c | c | c }
	
	\hline   & \textbf{Benign} & \textbf{Atypical} & \textbf{Malignant} & \textbf{Total \glspl{wsi}} & \textbf{Total Patients}   \\ 
	\hline Train &  $1460$ & $1011$ & $1186$ & $3657$  & $106$ \\
	\hline Validation &  $135$ & $90$ & $87$ & $312$  & $15$ \\
	\hline Test &  $242$ & $162$ & $166$ & $570$  & $30$ \\ 
	\hline
	\end{tabular}
	\caption{\gls{roi}-level split according to the lesion type.}
	\label{table:table5}
\end{table}

\begin{table} [H]
\centering
	\begin{tabular}{ l | c | c | c | c | c | c | c }
	
	\hline  & \textbf{N} & \textbf{PB} & \textbf{UDH} & \textbf{FEA} & \textbf{ADH} & \textbf{DCIS} & \textbf{IC}  \\ 
	\hline Train &  $357$ & $714$ & $389$ & $624$  & $387$  & $665$  & $521$\\
	\hline Validation &  $46$ & $43$ & $46$ & $49$  & $41$  & $40$  & $47$ \\
	\hline Test &  $81$ & $79$ & $82$ & $83$  & $79$  & $85$  & $81$ \\ 
	\hline
	\end{tabular}
	\caption{\gls{roi}-level split according to the lesion subtype.}
	\label{table:table6}
\end{table}

\section*{Data organization}

The \gls{bracs} dataset can be publicly accessed and downloaded via the \gls{bracs} website~\cite{bracs2021}. Anyone registering and agreeing with the terms of use (Creative Commons CC0 license) can freely download it. Once registered, the user can access the FTP server containing all the data. 

\begin{figure*}
      \centering
      \includegraphics[width=400pt]{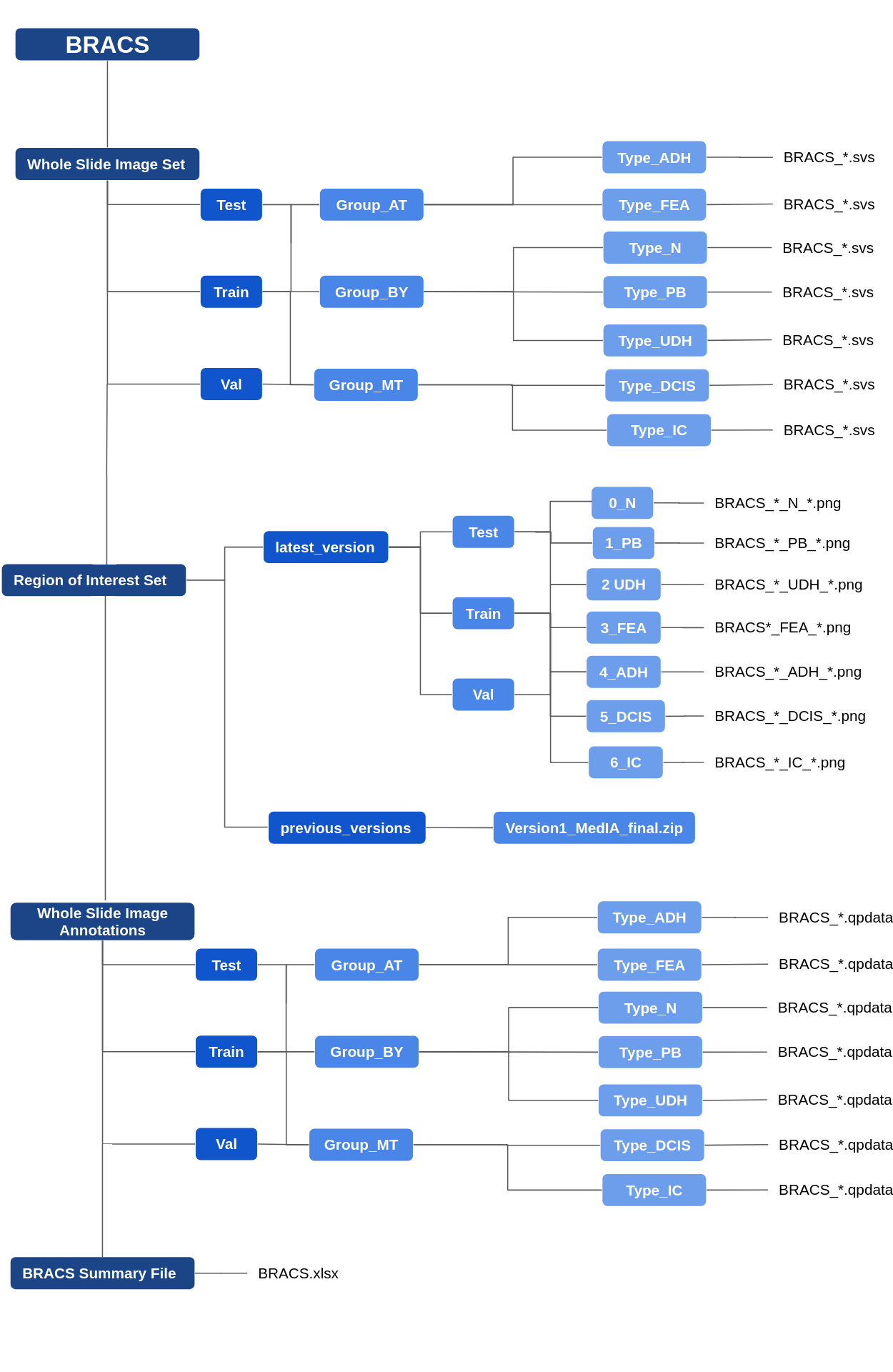}
      \caption{The organization of \gls{bracs} dataset folders.}
      \label{figure:struct}
\end{figure*}

The data are organized as follows.
The \glspl{wsi} are stored in the \textsl{Whole Slide Image Set} folder, that includes the train, validation, and test data. Each data split folder is further partitioned in Benign (BY), Atypical (AT) and Malignant (MT) folders, each of which includes folders corresponding to lesion subtypes. The \glspl{wsi} are stored as .svs files. All the files follow the same naming convention. For instance, the file \textsl{BRACS\_1238.svs} refers to the slide ID 1238, whose label is defined by the name of the folder that contains it. 
The \glspl{roi} are stored in the \textsl{Region of Interest Set} folder, which follows the same structure as the \gls{wsi} set. The files are stored in .png format, where the file \textsl{BRACS\_1238\_PB\_32.png} refers to the \gls{roi} number 32, extracted from the \gls{wsi} named \textsl{BRACS\_1238.svs}, and labeled as Pathological Benign.
The folder also includes a \textsl{previous\_versions} archive that contains a .zip file with data that have been used in a series of publications during the dataset collection process, \eg~\cite{pati20,pati21,jaume20,jaume21}.
The \gls{wsi} annotations are stored in the \textsl{The Whole Slide Image Annotations} folder, which follows the same structure as the \gls{wsi} set. It includes annotation files in .qpdata format (based on QuPath~\cite{bankhead2017}) for visualizing the \glspl{roi} inside their corresponding \gls{wsi}.
Finally, a summary file is provided as an .xlsx file, which reports for each \gls{wsi}, its label, reference set (training/validation/test), corresponding patient ID and the number of associated \glspl{roi}, if any. 
Figure~\ref{figure:struct} highlights the folder organization of \gls{bracs}. 


\clearpage
\bibliography{sample}

\section*{Acknowledgements} 

We would like to acknowledge Engr. Alessandro Manzoni from the IRCCS Fondazione Pascale for his support for management of technical instruments. We would like to acknowledge Ph.D Mario Sicuranza from ICAR-CNR for the development and technical implementation of the \gls{bracs} site.

\section*{Author contributions statement}
National Cancer Institute- IRCCS-Fondazione Pascale, Naples, Italy\\
Anna Maria Anniciello, Gerardo Botti, Maurizio Di Bonito, Florinda Feroce, Giosuè Scognamiglio\\
\\
Institute for High Performance Computing and Networking-CNR, Naples, Italy\\
Nadia Brancati, Giuseppe De Pietro, Maria Frucci, Daniel Riccio\\
\\
IBM Zurich Research Lab, Zurich, Switzerland\\
Pushpak Pati, Guillaume Jaume, Antonio Foncubierta, Maria Gabrani \\

\textbf{Contributions}\\
All authors have provided critical feedback during revision process and actively participated in preparation of the manuscript. \\
G.B., G.D.P and M.G. were responsible for the conceptualization of this project.\\
A.F and M.F. contributed to the conceptualization and were responsible for execution and management of this project.\\
M.D.B. was responsible for the institutional clinical databases and contributed to the annotation process.\\
A.M.A. and F.F. contributed to the clinical design of the dataset, data annotations and documentation for the project and organization of the validation of results.\\
P.P., N.B. and G.J. contributed to the
dataset organization and validation.\\
G.S. collected cases, extracted and scanned slides from their institutional clinical databases.\\
D.R. contributed to the conceptualization of this project.



\end{document}